\documentclass[twocolumn,twoside,slac_two]{revtex4}
\usepackage{graphicx}
\usepackage{fancyhdr}
\begin{document}
\title{Multiwavelength study of TeV Blazar Mrk421 during giant flare and observations of TeV AGNs with HAGAR}
\author{A. Shukla\footnote{amit@iiap.res.in},  P. R. Vishwanath, G. C. Anupama, T. P. Prabhu}
\affiliation{Indian Institute of Astrophysics, Bangalore, 560 034, INDIA}
\author{V. R. Chitnis, B. S. Acharya, R. J. Britto, B. B. Singh }
\affiliation{Tata Institute of Fundamental Research, Mumbai, 400 005, INDIA}
\author{P. Bhattacharjee, L. Saha}
\affiliation{Saha Institute of Nuclear Physics, Kolkata, 700 064, INDIA}

\begin{abstract}
The radiation mechanism of very high energy $\gamma$-ray emission from blazars and crucial parameters like magnetic field, and size of the emitting region are not well understood yet. To understand the above mentioned properties of blazars, we observed five nearby TeV $\gamma$-ray emitting blazars (Mrk421, Mrk501, 1ES2344+514, 1ES1218+304 and 3C454.3) and one radio galaxy (M87) using the High Altitude GAmma Ray (HAGAR) telescope. HAGAR is an array of seven telescopes located at Hanle, India to detect Cherenkov light caused by extensive air showers initiated by $\gamma$-rays. Mrk421 was observed to undergo one of its brightest flaring episodes on 2010 February 17, and detected by various experiments in X-rays and $\gamma$- rays. HAGAR observations of this source during 2010 February 13 - 19, in the energies above 250 GeV show an enhancement in the flux level, with a flux of 6-7 Crab units being detected on 2010 February 17. We present the spectral energy distribution of the source during this flaring episode. In addition to this, the analysis procedure to extract $\gamma$-ray signal from HAGAR data is discussed and preliminary results on all the AGNs are presented.

\end{abstract}

\maketitle

\thispagestyle{fancy}


\section{Introduction}
Blazars are a subclass of Active Galactic Nuclei (AGN) with a relativistic jet that is pointing in the direction of the Earth. They belong to the radio-loud AGNs and are characterized by a non-thermal spectrum extending up to high energies. Spectral energy distribution (SED) of high energy peaked TeV blazars show two broad peaks, one from infrared to X-ray energies and an other at X-ray to $\gamma$-ray energies. It is believed that first peak of the SED originates due to synchrotron radiation by relativistic electrons gyrating in the magnetic field of the jet plasma. The origin of high energy GeV/TeV peak is still under debate. This high energy peak could originate due to either inverse Compton (IC) scattering of synchrotron photons by same population of electrons which produces the synchrotron radiation (for a recent review of observations and models, see \cite{lab1}) or extremely energetic protons gyrating in strong magnetic field via synchrotron radiation \cite{lab2} or as IC and synchrotron emission from a proton-induced cascade \cite{lab3}. The blazar, Mrk421 (z=0.031), has been the first extragalactic source detected at $\gamma$-ray energy E $>$ 500 GeV \cite{lab4}. Using the newly commissioned High Altitude GAmma Ray (HAGAR) telescope system we have observed several TeV blazars in the last three years. Here, we report observations of Mrk421 in its high state of activity during February to April 2010 and  that of Mrk501 during March-June 2010. 
\section{HAGAR}
HAGAR, an array of Atmospheric Cherenkov Telescopes (ACT) using wave front sampling technique, is located at Hanle, at an altitude of 4300 meters AMSL, India. The main motivation behind setting up the $\gamma$-ray array at high altitude is to exploit the higher Cherenkov photon density and thus achieve lower energy threshold. HAGAR consists of an array of seven telescopes in the form of a hexagon, with one telescope at the center. All seven telescopes have seven para-axially mounted front coated mirrors of diameter 0.9 meter, with a UV sensitive photo-tube at the focus of individual mirror. Each telescope is separated by $50$ meters distance from its neighboring telescope. HAGAR Data Acquisition (DAQ) system is of CAMAC based. In addition, a parallel DAQ using commercial waveform digitizers (ACQIRIS make) is also used.\\
The performance of the HAGAR array has been studied by simulations, which are done in two steps; first step is generation of $\gamma$-ray and cosmic ray induced showers in atmosphere by using Monte Carlo simulation package CORSIKA, developed by KASKADE group \cite{lab5}. Second step is to study the response of the array towards the Cherenkov radiation produced by the simulated showers. The performance parameters such as energy threshold, collection area and sensitivity of the experiment are obtained by a detector simulation package indigenously developed by the HAGAR collaboration. Energy threshold of the HAGAR telescope is estimated as 204 GeV in case of vertically incident $\gamma$-ray showers for $\ge$ 4-Fold trigger condition for which the corresponding collection area is $3.2\times10^{8} cm^{2}$. HAGAR sensitivity is such that it will detect Crab nebula like source at a significance level of $5\sigma$ in 15 hours of observation \cite{lab6}. Cherenkov emission originated by induced air showers follows a spherical wavefront with a large radius of curvature and thickness of around 1 meter at observation level. This spherical wavefront is approximated as a plane wavefront in data analysis procedure, which is a good approximation at the observation level. The arrival direction of each shower in Cherenkov light pool is computed by measuring the relative arrival times of shower front at different telescopes. Normal to this plane front gives the arrival direction of the incident shower. The angle between direction of the shower axis and pointing direction of the telescope is defined as Space Angle ($\psi$). These Space Angles are constructed for every event by measuring relative arrival time difference at each telescope.\\
The observations were carried out by pointing all seven telescopes towards the source or background direction at a time. Each source run was followed by a background run with the same exposure time (typically  $40$ minutes) and covered the same zenith angle range as that of the source to ensure that observations are carried out at the same energy thresholds. Data selection was done by using parameters which characterize good quality data, in order to reduce systematic errors. Extraction of $\gamma$-ray signal was carried out by comparing on-source with off-source space angle distribution obtained during the same night at the same energy threshold. The excess events were computed from $0^{o}$ to Lower Limit (LL). LL is defined as the foot of distant half maximum point computed by fitting a Gaussian to the distribution for this analysis. Background space angle distribution was normalized with source space angle distribution by comparing the tails of the distributions (LL to $6.5^{o}$), since no $\gamma$-ray events are expected in this region. This normalization is required to compensate any possible change in observation conditions or change in sky conditions during the ON Source to OFF source.
\begin{figure}[t]
\centering
\includegraphics[width=60mm]{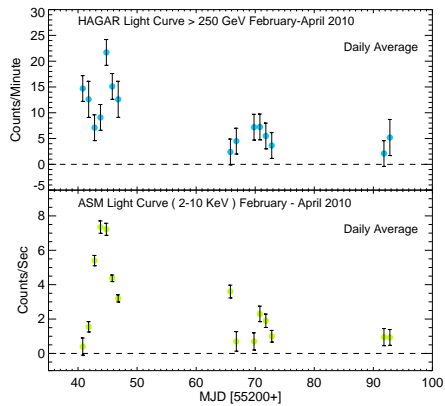}
\caption{The top panel shows daily average light curve of Mrk421 during the period of February - April 2010 in VHE $\gamma$-ray by HAGAR above 250 GeV and bottom panel shows the same in 2-10 keV X-rays by ASM. The X-axis should be read as 55200$+$.} \label{lc2010-f1}
\end{figure}
\begin{figure}[t]
\includegraphics[width=60mm]{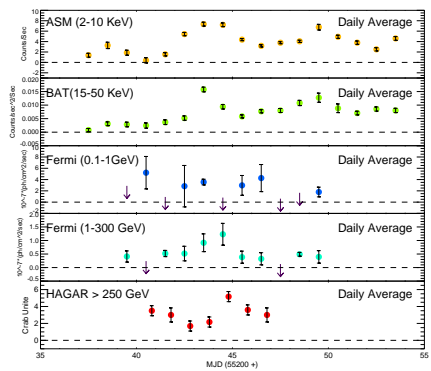}
\caption{The top four panels of are daily average of ASM, BAT, Fermi (100 MeV -1GeV) and Fermi (1-300 GeV) respectively and bottom panel in the plot corresponds to HAGAR data above 250 GeV during the month of February 2010 of Mrk421} 
\label{lc2010-f2}
\end{figure}
\begin{table*}[t]
\begin{center}
\caption{Results of observed blazars from HAGAR.}
\begin{tabular}{|l|c|c|c|c|}
\hline \textbf{Objects}&\textbf{Observation Period}&\textbf{Exposure Time (Hr)}&\textbf{Mean $\gamma$-ray Rate (per min.)}&\textbf{Significance}
\\
\hline Mrk421 & February 2010 & 8  & 13.4 $\pm$1.05& 12.7 \\ 
\hline Mrk421 & March- April 2010 & 11.2 & 4.3$\pm$0.85 & 5.1 \\
\hline Mrk501 & June 2010 & 5.6 & 5.73 $\pm$1.33& 4.3 \\
\hline
\end{tabular}
\label{l2ea4-t1}
\end{center}
\end{table*}
\section{Obervations}
\subsection{HAGAR}
Observations of Mrk421 in an active state were carried out for three months, during February-April 2010 on moonless nights by using HAGAR. Mrk421 was observed with a mean zenith angle of $6^{o}$. Observations of Mrk501 were also made during March-June, 2010 on moonless nights. The maximum zenith angle for Mrk501 observation during its observations were $25^{o}$.
\subsection{Fermi-LAT} 
The Large Area Telescope (LAT) is a pair production telescope \cite{lab7} on board Fermi spacecraft. LAT covers the energy range from 20 MeV to more than 300 GeV with field of view $\ge$ 2.5 sr. Fermi-Lat data\footnote{http://fermi.gsfc.nasa.gov/} 
from 12 - 22 February 2010 of Mrk421 above 100 MeV was analyzed with the standard analysis procedure (ScienceTools) provided by Fermi-Lat collaboration. A 10 degree region of interest (ROI) was chosen around the source for event reconstruction from the so-called ``diffuse'' event class data. To avoid the background of earth albedo we only retained events having a zenith angle $<105^{o}$. To determine the flux and spectrum of a source, the resulting data set was analyzed including Galactic and isotropic backgrounds with instrumental response function P6V3 DIFFUSE by using likelihood analysis (unbinned gtlike). We used a power law spectrum to model the source spectrum above the 100 MeV with integral flux and photon index as free parameters.
\subsection{PCA, ASM and BAT} 
Proportional Counters Array (PCA) is an array of five identical xenon filled proportional counter units (PCUs). The PCUs cover energy range from 2-60 keV with a total collecting area of 6500 cm$^{2}$. The archival X-ray data from PCA on board RXTE is analyzed for 17 February 2010 to obtain X-ray spectrum and light curve. We have analyzed Standard 2 PCA data which has a time resolution of 16 s with energy information in 128 channels. Data reduction was done with FTOOLS (version 5.3.1) distributed as part of HEASOFT (version 5.3). Data were filtered using standard procedure given in the RXTE Cook Book 4 for each of the observations. The background models were generated with the tool ``pcabackest'', based on RXTE GOF calibration files for a 'bright' source (more than 40 ct/sec/PCU).
The ``Dwell'' data from RXTE-ASM were obtained using ASM website\footnote{http://xte.mit.edu/} and these data were analyzed by the method discussed in \cite{lab8}. A daily average flux between 15-50 keV from Swift BAT is obtained from BAT website\footnote{http://heasarc.nasa.gov/docs/swift/results/transients/}.
\section{Results}
\subsection{Mrk421} 
We have analyzed data from Mrk421 collected by using HAGAR telescope during February - April, 2010.
Mrk421 was found to be in high state of activity during the entire period of HAGAR observations, and  was in its brightest state in February, 2010. Fluxes of $\gamma$-rays and X-rays decreased in later months, but were still higher than that during its quiescent state. Figure 1 contains the daily light curve of of Mrk421 in X-ray and $\gamma$-ray during the high state of activity. The upper panel shows the daily average of $\gamma$-ray flux. Bottom panel shows daily average in 2-10 keV from ASM on-board RXTE. It is clearly seen in the HAGAR as well as ASM light curves that source was in brightest state in the month of February, in both $\gamma$-rays and X-rays. All details of the HAGAR results are given in table 1.\\
We have further investigated the flaring behavior in the month of February 2010. HAGAR telescope has detected Mrk421 in a high state of very high energy (VHE) $\gamma$-ray flux above 250 GeV, during the observations of 13-19 February, 2011. The maximum flux above 250 GeV is found to be between 6-7 Crab units (1 Crab unit $\sim$ 4.2 $\gamma$-rays/minute above 250 GeV) on 2010 February, 17.
\subsubsection{Flux Variability}
A quasi-simultaneous light curve is obtained of Mrk421 in X-rays and $\gamma$-rays band by using archived data of soft X-ray by ASM on board RXTE, hard X-ray data from BAT on board Swift and $\gamma$-ray data from LAT on board Fermi with the observed HAGAR ($>$250 GeV) data for month of February 2010 with a one day binning. The light curve is shown in Figure 2. The top 4 panels correspond to ASM, BAT, Fermi-LAT (0.1-1 GeV) and Fermi-LAT (1-300 GeV) respectively. The bottom panel in the plot corresponds to HAGAR data above 250 GeV. A clear variation in flux over a period of seven days in X-ray as well as $\gamma$-rays were observed. The peak flux in X-rays and low energy $\gamma$-rays by ASM, BAT and Fermi-LAT were detected on 16 February, 2010 but GeV/TeV gamma ray flux reached peak with a lag of one day on 17 February, 2011.\\
The observed X-ray variability could be explained by the synchrotron self-Compton (SSC) cooling mechanism. BAT light curve showed faster variability than ASM which could be due to the cooling effect of high energy electrons, which produce X-rays at 15-50 keV range.
The observed $\gamma$-ray variability could be mainly divided into two bands, $<$ 1 GeV and above 1 GeV. $\gamma$-rays (0.1-1 GeV) observed by LAT are produced by low energy electrons by IC scattering of a UV synchrotron photons. LAT detected a significant variation in the flux in $0.1-1 GeV $ band observed over the period of eleven days. The observed $>$ 1 GeV VHE $\gamma$-rays by Fermi-LAT and HAGAR could be produced by IC scattering of the electron having Lorentz factor $\sim 10^{4}-10^{5}$ with X-ray photons produced by tail of the electron energy distribution.\\
\subsubsection{Intra-Day variability}
Fermi-LAT data showed an intra-night flux as well as spectral variability during the TeV flare on 17 February, 2010. It has shown a flaring behavior in first 10 hours of LAT observations and then become quiescent for next few hours and start flaring up again later in the night. A similar trend was detected by the VERITAS collaboration \cite{lab9}. HAGAR also observed a continuous decrease in the flux over a period of $\sim$ 4 hrs, which were simultaneous with LAT during later part of the night.
\subsubsection{Spectral Energy Distribution}
 We attempted to obtain the SED of Mrk421 during the flare of 17 February 2011 by using soft X-ray data from PCA on board RXTE, $\gamma$-ray data from LAT on board Fermi and HAGAR data.
Spectral analysis of X-ray data was done by using xspec, PCA spectral data of 17 February was fitted with a cutoff powerlaw with line of sight absorption. The line of sight absorption was fixed to neutral hydrogen column density at $1.38\times10^{2} cm^{-2}$ \cite{lab10}. The Fermi-LAT data is divided into four bins (0.1-1 GeV, 1-3 GeV, 3-10 GeV and 10-300 GeV) to obtain spectrum of Mrk421 on 17 February by freezing photon index to 1.39 which is obtained by taking 0.1 -300 GeV data. 
 \begin{figure}[!t]
  \vspace{5mm}
  \centering
  \includegraphics[width=70 mm]{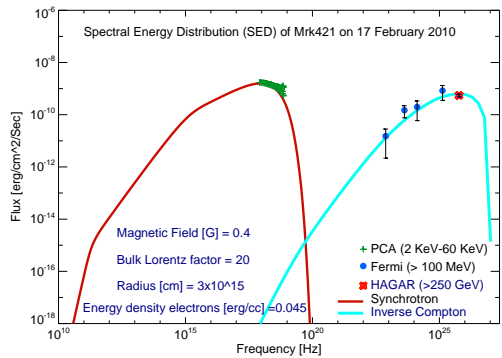}
  \caption{Spectral Energy Distribution of Mrk421 of 17 February, 2010.}
  \label{simp_fig}
 \end{figure}
A one zone homogeneous SSC model \cite{lab11} is fitted to the data to obtain the SED. This model assumes a spherical blob of radius R and uniform magnetic field B, moving with respect to the observer with the Lorentz Factor $\Gamma$ and  is filled with a homogeneous non-thermal electron population. We used equipartition of the fields to compute the best fit parameters. A best fit of SED is obtained for the parameters given below:\\
Bulk Lorentz factor of emitting blob ($\Gamma$) : 20 \\
Strength of magnetic field in the jet frame (B) : 0.4 Gauss \\
Comoving radius of emission volume : $3\times10^{15}$  cm \\
Jet-frame energy density of the electrons (U) : 0.045 erg/cc   
\subsection{Mrk501 and other Blazars}
HAGAR detected TeV blazar Mrk501 in relatively high state of activity during June 2010. The preliminary analysis of Mrk501 data indicates an average flux of the source during June 2010 was $1.36$ Crab units. Data analysis is going on for all the TeV blzaras.
\begin{acknowledgments}
This work used results provided by the ASM/RXTE teams at MIT. This research has
also made use of data obtained from the High Energy Astrophysics Science Archive Research Center (HEASARC), provided by NASA's Goddard Space Flight Center. We are grateful to NASA and engineering and technical staff of IIA and TIFR who have taken part in the construction of the telescopes and contributed to setting up of the front-end electronics and the data acquisition. 
\end{acknowledgments}
\bigskip 

\end{document}